\documentstyle[preprint,aps]{revtex}
\tightenlines

\begin{document}
\newcommand{\beq}{\begin{equation}}
\newcommand{\eeq}{\end{equation}}
\newcommand{\beqa}{\begin{eqnarray}}
\newcommand{\eeqa}{\end{eqnarray}}
\newcommand{\sr}{\sqrt}
\newcommand{\fr}{\frac}
\newcommand{\mn}{\mu \nu}
\newcommand{\G}{\Gamma}

\draft
\preprint{ INJE-TP-01-03, hep-th/0103241}
\title{ Standard cosmology from the brane cosmology
with a localized matter }
\author{  Y.S. Myung\footnote{E-mail address:
ysmyung@physics.inje.ac.kr} }
\address{Relativity Research Center and School of Computer Aided
School\\ Inje University,
Gimhae 621-749, Korea}
\maketitle

\begin{abstract}

We discuss  the brane cosmology in the 5D anti de Sitter
Reissner-Nordstrom (AdSRN$_5$) spacetime. A brane with the
tension $\sigma$ is defined as the edge of an AdSRN$_5$ space with mass $M$
and charge $Q$.
In this case we get the CFT-radiation term $(\rho_{CFT})$ from $M$ and the
charged dust $(-\rho^2_{cd})$ from $Q^2$ in the Friedmann-like equation.
 However, this equation is  not
justified because it contains   $\rho_{cd}^2$-term with
the negative sign. This is unconventional in view of the standard cosmology.
In order to resolve this problem, we introduce
a localized dust matter which satisfies $P_{dm}=0$. If
$\rho_{dm}=\fr{\sqrt 3}{2}\rho_{cd}$,
the unwanted $-\rho_{cd}^2$ is cancelled against $\rho_{dm}^2$
and
thus one recovers a standard Friedmann-Robertson-Walker universe with
CFT-radiation and dust matter.
For  the stiff matter consideration, we can set $\rho_{csti}\sim Q^2$
with the negative sign. Here we introduce a massless
scalar which plays  the role of a stiff matter with
$P_{sca}=\rho_{sca}$ to cancel  $-\rho_{csti}$.
In this case, however, we find a mixed version of the standard and brane cosmologies.

\end{abstract}

\newpage

Recently Verlinde has introduced entropy bounds  to establish
  the holographic principle in the  Friedmann-Robertson-Walker (FRW) universe\cite{VER}.
 Representing  a radiation
as   a conformal field theory (CFT) at high temperature
 within a 5D anti de Sitter Schwarzschild (AdSS$_5$)-bulk
theory, the Cardy-Verlinde's  formula maps to Friedmann equation\cite{BFV}.
This means that the Friedmann
equation keeps information about  thermodynamic relations  of the CFT.
Initially the Friedmann equation
 has nothing to do with the thermodynamic Cardy-Verline's formula.
In order to understand this connection,
we   note that in this approach,
the  AdS/CFT correspondence plays the role of an important tool to realize
the holographic principle.

Within the brane cosmology context, there exist two distinct terms in the Friedmann equation
in compared with
the standard cosmology: one is  $\rho^2$-term\footnote{Sometimes, this term
is awkward to interpret and sometimes it is useful for resolving the old issues
of the standard cosmology\cite{BDL,KRA,HMR}.} due to the localized matter distribution
and the other is the presence of the non-local term arisen  from
the bulk configuration. Even though the localized matter is absent,
an observer on the brane  finds  the  radiation-like matter which comes from
 the non-local term. This holographic term is interpreted as the above CFT-radiation. Further
the  entropy and temperature can be
expressed solely in terms of the Hubble parameter $H=\dot
a/a$\cite{SV} when the brane crosses the
horizon $r=r_+$ of the AdSS$_5$ black hole.
 In this case the brane is considered as the  edge of an AdSS$_5$ space.

Let us  discuss this situation more explicitly. The brane
starts at $r=0$ (big bang) inside the small black hole ($\ell>r_+$),
crosses the horizon at $a=r_+$, and
expands until it reaches  maximum size $a=r_m$. And then the brane
contracts and  it falls the black hole again. Finally the brane  disappears (big crunch). A
bulk observer in  AdSS$_5$-space finds two interesting moments (two points in
the Penrose diagram\cite{MSM}) when the brane crosses the past (future) event
horizons. Authors in \cite{SV} showed that at these times the
Friedmann equation controlling the dynamics of the brane
coincides with the Cardy-Verlinde's formula describing the entropy-energy relation
 of the CFT defined on the brane. The location of the  horizon  corresponds to the
holographic point. Hence this
 event can be interpreted as a consequence of the cosmological
 holography because one used mainly  the  AdS/CFT correspondence.

More recently, authors in \cite{BMU} studied  a similar case  but in
different background of the  Reissner-Nordstrom black hole
\footnote{The approach to this direction was first made in ref.\cite{BV}.
In that paper, authors mentioned the charge term in the Friedmann
equation as the stiff matter even though with a negative sign.
However, in this  work, we focus on how to avoid the negative term
to obtain the standard cosmology in the AdSRN$_5$ background.
Hence our viewpoint is basically  different from that of \cite{BV}.}.
Considering the
brane-universe  in this charged background, one may get either
the  CFT-radiation matter ($\rho_{CFT}=E/V=M\ell/aV$) or
the charged-dust matter ($\rho_{cd}=Q/V$).
The presence of the  bulk black hole charge gives rise to
$-\rho_{cd}^2$-term which is just a characteristic  of the brane
cosmology with a localized matter distribution.
Unfortunately, one finds here the negative sign  in front of  $\rho_{cd}^2$ in the
Friedmann-like equation. This negative quadratic term is obviously
unconventional in terms of cosmology
because the standard cosmology does not include such a term
\footnote{Assuming the equation of state $p=\omega \rho$, the conservation law
gives rise to $\rho \sim a(\tau)^{-3(1+\omega)} $ in four dimensions.
Here the causality requires $|\omega|\le 1$. $\omega=0, +1/3, +1,-1$
correspond to nonrelativistic matter (dust), relativistic matter (radiation), stiff
matter, cosmological constant, respectively. Even if the quintessence is concerned,
one requires $-1<\omega<-1/3$.  Hence it is unnatural to include
a negative term of $-\rho_{cd}^2 \sim -a^{-6}$ for a cosmological evolution
 in the Friedmann equation
except the negative cosmological constant of  $\omega=-1$.}.
This implies that mass $M$ and charge $Q$
of the Reissner-Nordstrom black hole affect the
brane moving in this black hole background differently.

In this paper, we concentrate on resolving
 this embedding problem  of the moving domain wall (brane)
into  an AdSRN$_5$-black hole spacetime.
 This can  be done   by introducing any localized dust matter $\rho_{dm}$ with $P_{dm}=0$.
It turns out that if $\rho_{dm}= \fr{\sqrt{3}}{2}\rho_{cd}$,
the unwanted-term $-\rho_{cd}^2$ is
cancelled against $\rho_{dm}^2$.
 Hence we  can recover   a Friedmann equation for the standard cosmology
which is composed of the  CFT-radiation matter and dust matter.
On the other hand, following the stiff matter interpretation of  the
charge\cite{BMU,BV}, one has $-\rho_{csti} \sim -Q^2/V^2$.
In this case, to cancel $-\rho_{csti}$ we have
to introduce a massless scalar without the potential
which is known as  a stiff matter with equation of state $P_{sca}=\rho_{sca}$.
 However,  the resulting Friedmann
 equation contains the standard as well as  brane cosmological
 features.

For a cosmological embedding, let us start with an
AdSRN$_5$-spacetime\cite{BMU} ,
\beq
ds^{2}_{5}= g_{MN}dx^Mdx^N= -h(r)dt^2 +\fr{1}{h(r)}dr^2 +r^2
\left[d\chi^2 +f_{k}(\chi)^2(d\theta^2+ \sin^2 \theta d\phi^2)
\right],
\label{BMT}
\eeq
where $k=0,\pm1$. $h(r)$ and $f_k(\chi)$ are given by
\beq
h(r)=k-\fr{m}{r^2}+\fr{q^2}{r^4}+ \fr{ r^2}{\ell^2},~~~
f_{0}(\chi) =\chi, ~f_{1}(\chi) =\sin \chi, ~f_{-1}(\chi) =\sinh
\chi.
\eeq
In the case of $q=0$, we have an uncharged AdSS$_5$-space. The
extremal black hole with $m/2=|q|$ is not allowed in this context\cite{Cai}.
 Using the moving domain wall approach\cite{KRA}, we  can derive
the 4D induced line element from Eq.(\ref{BMT})
\beqa
ds^{2}_{4}&&=-d \tau^2 +a(\tau)^2
\left[d\chi^2 +f_{k}(\chi)^2(d\theta^2+ \sin^2 \theta d\phi^2)
\right]
\nonumber \\
&&\equiv h_{\mu \nu}dx^{\mu} dx^{\nu},
\label{INM}
\eeqa
where we use the Greek indices only  for the brane.
Actually the embedding of the moving domain wall into an AdSRN$_5$ space  is a
$2(t,r) \to 1(\tau)$-mapping: $t \to t(\tau), r \to a(\tau)$.
Here the scale factor $a$ will be determined
by the Israel junction condition\cite{ISR}. The extrinsic curvature is given by

\beqa
&&K_{\tau\tau}=K_{MN} u^M u^N =(h(a) \dot t)^{-1}(\ddot a +h'(a)
 /2)=\fr{\ddot a +h'(a)/2}
{\sqrt{\dot a^2 +h(a)}}, \\
&&K_{\chi\chi} = K_{\theta\theta}=K_{\phi\phi}
=- h(a) \dot t a=-\sqrt{\dot a^2 +h(a)}~a,
\eeqa
where  prime stands for  derivative with respect to $a$.
The presence of any localized matter on the brane  including the brane tension
 implies
that the extrinsic curvature jumps  across the brane.
This jump is  described  by the Israel junction
condition\footnote{The bulk Maxwell term and the Hawking-Ross term are necessary for
embedding of the brane in the
charged black hole background. However, the flux across the brane does not
vary and the junction condition remains unchanged because the charge $q$ is
fixed\cite{GP}. Furthermore we consider  only one-sided brane
cosmology to see the moving domain wall picture clearly. Hence the Z$_2$-symmetry which was
usually issued  in two-sided brane cosmology is no longer
considered here.}

\beq
K_{\mu \nu}=-\kappa^2 \left(
T_{\mu\nu}-\fr{1}{3}T^{\lambda}_{\lambda}h_{\mu\nu} \right)
\label{4DI}
\eeq
with $\kappa^2=8 \pi G_5^N$. For  cosmological purpose
we  introduce the 4D perfect fluid  as a localized stress-energy tensor on the
brane
\beq
T^{b+m}_{\mu \nu}=(\varrho +p)u_{\mu}u_{\nu}+p\:h_{\mu\nu}.
\label{MAT}
\eeq
Here $\varrho=\rho+ \sigma$ $(p=P-\sigma)$, where $\rho $ $(P)$ is the energy density (pressure)
of the localized matter and $\sigma$ is the brane tension.
In the case of $\rho=P=0$, the r.h.s. of
Eq.(\ref{4DI}) leads to  a form of the RS case as $-\fr{\sigma \kappa^2}{3}
h_{\mu\nu}$.
From Eq.(\ref{4DI}), one finds
 the space component of the junction condition

\beq
\sqrt{h(a) + \dot a^2}=\fr{\kappa^2}{3}\sigma a.
\label{SEE}
\eeq
For a single AdSRN$_5$ spacetime,
we have the fine-tuned brane tension $\sigma=3/(\kappa^2\ell)$.
The above equation  then  leads to
\beq
H^2=- \fr{k}{a^2} +\fr{m}{a^4} -\fr{q^2}{a^6},
\label{HHH}
\eeq
where $H=\dot a/a$ is the Hubble parameter. The non-local term of $m/a^4$
originates from the electric (Coulomb) part of the bulk Weyl tensor, $E_{00} \sim
m/r^4$\cite{SMS,MSM}. This term behaves like radiation for either dark matter\cite{KRA} or
CFT \cite{BFV}.
Especially for a closed brane  ($k=1$),
 we have  $m=\omega_4 M$ and
$q^2=\fr{3 \omega^2_4}{16}Q^2$ with $\omega_4= \fr{16 \pi G_5^N}{3
V(S^3)}$ and $V=a^3 V(S^3)$. Using the AdS/CFT correspondence, we obtain the
bulk-boundary relations:
$M=\fr{a}{\ell}E, G_5^N=\fr{\ell}{2} G_4^N$. Here $M(Q)$ is the ADM mass (charge) which
are measured at the spatial infinity of AdS$_5$ space.
 Then one finds a universe filled with the  CFT-radiation
 and  charged-dust
\beq
H^2=- \fr{1}{a^2} +\fr{8\pi
G_4^N}{3}\rho_{CFT}-\fr{\kappa^4}{12}\rho_{cd}^2
,~~~\rho_{CFT}=\fr{E}{V},~~\rho_{cd}=\fr{Q}{V},
\label{CRA}
\eeq
where $\rho_{CFT}$ scales like $a^{-4}$, whereas $\rho_{cd}^2$ behaves like
$a^{-6}$. Hence the charge density ($\rho_{cd} \sim a^{-3}$) plays the same role of a dust
matter.
It seems  that the Friedmann-like equation (\ref{CRA}) is awkward
to be interpreted as a cosmological one
because the last term is a negative one. Actually a negative energy density square is not
allowed because  the standard cosmology does not include such a term.
A way to resolve this problem is to cancel the last term by introducing an appropriate
localized matter. Because the last belongs to a dust matter,
we have to introduce the same kind of a localized  matter with the energy $E_{dm}$.
In this case Eq.(\ref{MAT}) takes the form
\beq
T^{b+dm}_{\mu \nu}=(\varrho
+p)u_{\mu}u_{\nu}+p\:h_{\mu\nu}.
\label{RMAT}
\eeq
Here $\varrho=\rho_{dm}+\sigma (\rho_{dm}=E_{dm}/V)$ and $p= -\sigma$ with the
dust equation of state: $P_{dm}=0$.
Using  the conservation law ($\dot \varrho + 3H (\varrho +p)=0$),
 we can easily check $\rho_{dm} \sim
a^{-3}$. Further Eq.(\ref{SEE}) is given by
\beq
\sqrt{h(a) + \dot a^2}=\fr{\kappa^2}{3}(\rho_{dm}+\sigma) a
\label{SE2}
\eeq
which leads with $\sigma=3/\kappa^2\ell$ to
\beq
H^2=- \fr{1}{a^2} +\fr{8\pi
G_4^N}{3}(\rho_{CFT}+\rho_{dm})+
\kappa^4\big(\fr{\rho_{dm}^2}{9}-\fr{\rho_{cd}^2}{4}\big).
\label{CR3}
\eeq
If $\rho_{dm}=\fr{\sqrt{3}}{2}\rho_{cd}$, the unwanted
$\rho_{cd}^2$-term disappears and one finds the Friedmann equation for
a standard cosmology with a definitely positive energy density
($\rho_{eff}>0$)
\beq
H^2=- \fr{1}{a^2} +\fr{8\pi
G_4^N}{3}\rho_{eff},~~~\rho_{eff}\equiv \fr{E_{eff}}{V}=\rho_{CFT}+ \rho_{dm},
~~~E_{eff}=E+E_{dm}.
\label{CR4}
\eeq
This is our main result. The effective density is composed of
two different matters:  CFT-radiation matter ($E \sim 1/a)$ and  dust matter
($E_{dm}=$ constant).
Here we do not expect the circular diagram when the Friedmann
equation (\ref{CR4}) can be expressed as the relation among the
Bekenstein entropy $S_B= \fr{2 \pi a}{n} E_{eff}$, the
Bekenstein-Hawking entropy $S_{BH}= (n-1) \fr{V}{4 G_4^N a}$, and
the Hubble entropy $S_{H}= (n-1) \fr{HV}{4 G_4^N }$
\beq
S_H^2 + (S_B-S_{BH})^2=S_B^2.
\label{ENT1}
\eeq
This is so because $S_B$ does not remain constant during the
cosmological evolution. In the case of the CFT-radiation matter
only, the Bekenstein entropy is constant with respect to the cosmic
time $\tau$\cite{VER}.

On the other hand, we may consider  the charge term as a stiff matter\cite{BMU,BV,Youm1}.
In this case, Eq.(\ref{CRA}) is rewritten as
\beq
H^2=- \fr{1}{a^2} +\fr{8\pi
G_4^N}{3}\tilde \rho_{eff},~~~\tilde \rho_{eff}=\rho_{CFT}-\rho_{csti}
\label{CRA2}
\eeq
with
\beq
\rho_{csti}=\fr{1}{2} \Phi \rho_Q,~ \Phi\equiv \phi \fr{\ell}{a}=
 \fr{3}{8} \fr{\omega_3 Q
\ell}{a^3},~\rho_Q=\fr{Q}{V}.
\eeq

Here $\phi$ is the bulk electrostatic potential which is difference between
the horizon and infinity, while $\Phi$ is its boundary CFT-potential.
$\rho_Q$ is the R-charge density of the CFT. So the cosmological evolution
will be driven by the the CFT-radiation energy $E$ and the electric potential
energy $\fr{1}{2} \Phi Q$ due to the R-charge $Q$.
Introducing the R-charge entropy of $S_Q= \fr{2 \pi a}{n} \cdot \fr{1}{2} \Phi
Q$, the entropy-relation is given by
\beq
S_H^2 + (S_B-S_Q-S_{BH})^2=(S_B-S_Q)^2.
\label{ENT2}
\eeq
In this case we note that $S_B-S_Q$ does not remain constant
during the evolution. Hence one cannot expect the circular diagram
for the CFT-radiation matter.
Obviously the above equation induces some problem to interpret it
as a cosmological evolution equation in Minkowskian  closed
universe. This is so because if $\rho_{CFT}<\rho_{csti}$,
$\tilde \rho_{eff}$ gives rise to a negative energy density.
This is the case that happens at the very early universe when
 $\rho_{csti} \sim 1/a^6$ is  more dominant than $\rho_{CFT}\sim
 1/a^4$.
This is not allowed for the standard  cosmology\footnote{In Mirage cosmology, however,
$\rho$ and $P$ may not stay positive during the expansion\cite{Youm}.}.
 Hence we would like  to resolve this problem by
introducing
an appropriate matter.
 One
candidate for stiff matters is a massless scalar without the potential
whose stress-energy tensor is given by
\cite{Youm,SI}
\beq
T_{\mu \nu}^{sca}=\partial_\mu \varphi \partial_\nu \varphi
- \fr{1}{2}(\partial \varphi)^2 h_{\mu\nu}.
\label{SMAT}
\eeq
If we assume $\varphi=\varphi(\tau)$ for cosmological purpose, this leads to
$T^\mu_\nu= {\rm diag}[-\dot \varphi^2/2,\dot \varphi^2/2,\dot \varphi^2/2,\dot
\varphi^2/2]\equiv {\rm diag}[-\rho_{sca},P_{sca},P_{sca},P_{sca}]$
 which satisfies obviously the stiff equation of state as
 $P_{sca}=\rho_{sca}$. In this case  we easily get $\rho_{sca} \sim a^{-6}$
 from the conservation  law.
Considering the  brane tension and scalar matter separately,
  Eq.(\ref{SEE}) takes the form
 \beq
\sqrt{h(a) + \dot a^2}=\fr{\kappa^2}{3}(\rho_{sca}+\sigma) a.
\label{SE1}
\eeq
This  leads to
\beq
H^2=- \fr{1}{a^2} +\fr{8\pi
G_4^N}{3}(\rho_{CFT}-\rho_{csti}+\rho_{sca})+
\kappa^4 \fr{\rho_{sca}^2}{9}.
\label{CR1}
\eeq
If $\rho_{csti}=\rho_{sca}$, the unwanted
$\rho_{csti}$-term disappears and one finds Friedmann-equation for
a mixed version of  the standard and brane cosmologies
\beq
H^2=- \fr{1}{a^2} +\fr{8\pi
G_4^N}{3}\rho_{CFT}+ \kappa^4\fr{\rho_{sca}^2}{9},~~~
\rho_{sca}=\fr{\dot \phi^2}{2}.
\label{CR2}
\eeq

 An embedding of the moving domain wall (brane) into the  anti de
Sitter-charged black hole spacetime  gives rise to the
negative electric energy density in   the Friedmann equation.
It is shown that  this problem can be resolved
 by choosing an  appropriate  stiff matter.
Even though we succeed in obtaining the positive energy density,
this action gives rises to the high-order term of $\rho_{sca}^2 \sim a^{-12}$
which have ever not been found in the standard and brane cosmology.
Such a high-order term, if it exists,
will contribute to the very early evolution of the universe significantly.

Finally we summarize our results.
Our first view is to regard the charge term ($-Q^2/a^6)$ as $-\rho_{cd}^2\sim
-Q^2/V^2$. Reminding that the brane cosmology automatically
produces  $+\rho^2$-term, we  resolve it by introducing
 a dust matter with its equation of state $P_{dm}=0$.
As a result, we obtain a standard FRW universe filled with two different matters:
 CFT-radiation matter  and dust matter.

The second view is to consider the charge term as a stiff
matter\cite{BMU}. In this case, we have $-\rho_{csti}\sim -Q^2/a^6$
which induces the  negative energy  problem in cosmology.
To resolve this, we  introduce the same kind of matter on the brane.
One candidate is just a massless scalar that is known to satisfy
the stiff equation of state $P_{dm}=\rho$. However, a resultant
equation is a mixed Friedmann equation for standard and brane
cosmologies. In this sense, our first view is more promising
than the second one in obtaining the standard Friedmann equation from
the brane cosmological picture.
Although we obtain the entropy-relations for the two cases, these
do not belong to the circular diagram. Hence we conclude that the charge $Q$ of the
Reisser-Norstrom black hole plays a different role from the mass $M$ of the black
hole. Two are holographic non-local matters which come from the bulk charged
black hole. But these give rise to different matters : one provides
the CFT-radiation matter and the other gives either negative dust
matter or negative stiff matter. Also we find that the
circular entropy-relation for pure CFT-radiation which is derived from
the uncharged black hole  does not holds
for the charged black hole background.

 Further, it is  interesting
to study the connection between the Friedmann equations
(\ref{CR4}) and (\ref{CRA2}) and the Cardy-Verlinde's formula
for the entropy-energy relation when the brane crosses the horizon of AdSRN$_5$ black hole.
It was shown that this connection holds even for the charged black
hole\cite{Youm1}.

Finally we wish to remark that authors in\cite{CEG} considered a
similar issue  and derived a Friedmann equation\footnote{I thank to
C. Grojean for pointing  out this} . Also within  the Randall-Sundrum
context, cancellations  between
the bulk and the brane to recover
 standard cosmology were first proposed in\cite{CGK}.

\section*{Acknowledgments}

This work was supported in part by the Brain Korea 21 Program, Ministry of
Education, Project No. D-0025 and KOSEF, Project No. R02-2002-000-00028-0.


\begin{thebibliography}{99}
\bibitem{VER} E. Verlinde, hep-th/0008140.


\bibitem{BFV} R. Brustein, S. Foffa
and G. Veneziano, hep-th/0101083; D. Klemm, A. Petkou and G.
Siopsis, hep-th/0101076; B. Wang, E. Abdalla and R. K. Su,
hep-th/0101073; S. Nojiri and S. Odintsov, hep-th/0011115; F.L.
Lin, hep-th/0010127; D. Kutasov and F. Larsen, hep-th/0009244;
Y.S. Myung, hep-th/0102184;
D. Birmingham and S. Mokhtari, hep-th/0103108;
S. Nojiri and S. Odintsov, hep-th/0103078;
L. Anchordoqui, C. Nunez, K. Olsen, hep-th/0007064.

\bibitem{SV} I. Savonije and E. Verlinde, hep-th/0102042.

\bibitem{MSM}  S. Mukhoyama, T. Shiromizu and K. Maeda, hep-th/9912287.


\bibitem{BMU} A. K. Biswas and S. Mukherji, hep-th/0102138.

\bibitem{BV} C. Barcelo and M. Visser, hep-th/0004056.

\bibitem{BDL} P. Binetruy, C. Deffayet and D. Langlois,
Nucl. Phys. B 565 (2000) 269 [hep-th/9905012];
 P. Binetruy, C. Deffayet, U. Ellwanger and D. Langlois, Phys. Lett. B 477
(2000) [hep-th/9910219].


\bibitem{KRA} P. Kraus, JHEP 9912 (1999) 011 [hep-th/9910149];
 D.  Ida, JHEP 0009 (2000) 014;
   H. Collins and B. Holdom, hep-ph/0003173;
  H. Stoica, S. H. H.  Tye and I. Wasserman, hep-th/0004126;
 N. J. Kim, H. W. Lee, and Y. S. Myung,
hep-th/0101091.

\bibitem{HMR}A. Davis, I. Vernon, S. Davis, and W. Perkins, hep-ph/0008132;
 A. Hebecker and J. March-Russell, hep-ph/0103214.

\bibitem{Cai}A. Chamblin, R. Emparan, C. Johnson, and R. Myers, hep-th/9902170;
 R. G. Cai, hep-th/0102113.




\bibitem{ISR}  W. Israel, Nuovo Cim. B44 (1966) 1; {\it ibid}.
B48 (1967) 463.

\bibitem{GP} J. P. Gregory and A. Padilla, hep-th/0204218.

\bibitem{SMS} T. Shiromizu, K. Maeda, and M. Sasaki,
gr-qc/9910076.

\bibitem{Youm} D. N. Youm, hep-th/0011290.
\bibitem{SI} T. Shiromizu and D. Ida, hep-th/0102035.
\bibitem{CEG} C. Csaki, J. Erlich, and C. Grojean, hep-th/0012143.
\bibitem{CGK} C. Csaki, M. Graesser, C. Kolda, and J. Terning, heh-ph/9906513;
J. Cline, C. Grojean, and  G. Servant, hep-ph/9906523.

\bibitem{Youm1} D. N. Youm, hep-th/0105249.

\end{thebibliography}
\end{document}